\documentclass[twocolumn,aps,prd,superscriptaddress,nofootinbib,preprintnumbers,floatfix]{revtex4-2}

\usepackage{graphicx}
\usepackage{multirow}
\usepackage[colorlinks=true,citecolor=blue,urlcolor=blue,linkcolor=blue]{hyperref}

\usepackage{amssymb}
\usepackage{amsmath}
\usepackage{wasysym,xcolor}
\usepackage{xspace}
\usepackage{tabularx}

\newcommand{\nn}{\nonumber}
\newcommand{\beq}{\begin{equation}}
\newcommand{\eeq}{\end{equation}}
\newcommand{\beqa}{\begin{eqnarray}}
\newcommand{\eeqa}{\end{eqnarray}}

\newcommand{\GeV}{{\rm GeV}}

\def\lqcd{\Lambda_{\rm QCD}}
\def\d{{\rm d}}

\def\hqs{\ensuremath{\hat q^2}}
\def\rt{\ensuremath{\rho_\tau}}
\def\hmts{\rt}

\newcommand{\Bbar}{\,\overline{\!B}{}}
\newcommand{\Dbar}{\,\overline{\!D}{}}
\newcommand{\Kbar}{\,\overline{\!K}{}}
\def\B0bar{\Bbar{}^0}
\def\D0bar{\Dbar{}^0}
\def\K0bar{\Kbar{}^0}

\newcommand{\pBPA}{PA-$B$\xspace}
\newcommand{\pNPA}{PA-$\bar\nu$\xspace}

\def\OMIT#1{{}}

\makeatletter
\g@addto@macro\bfseries{\boldmath}
\let\Hy@backout\@gobble
\makeatother

\begin{document}

\preprint{CALT-TH-2023-003}

\title{\texorpdfstring{$B\to X\tau\bar\nu$}{B->X tau nu}: spinning the \texorpdfstring{$\tau$}{tau} polarization axis}

\title{Polarizing the \texorpdfstring{$\tau$}{tau} in \texorpdfstring{$B\to X\tau\bar\nu$}{B->X tau nu}: a pain in the axis}

\title{Spatial orientation matters: \texorpdfstring{$\tau$}{tau} polarization in \texorpdfstring{$B\to X\tau\bar\nu$}{B->X tau nu} along different axes  }

\title{Juggling the \texorpdfstring{$\tau$}{tau} polarization axes in \texorpdfstring{$B\to X\tau\bar\nu$}{B->X tau nu}}

\title{Give 'em the axes: \texorpdfstring{$\tau$}{tau} polarization in \texorpdfstring{$B\to X\tau\bar\nu$}{B->X tau nu}}

\title{Sharpening the \texorpdfstring{$\tau$}{tau} polarization axes in \texorpdfstring{$B\to X\tau\bar\nu$}{B->X tau nu}}

\title{Choosing different axes for the  \texorpdfstring{$\tau$}{tau} polarization in \texorpdfstring{$B\to X\tau\bar\nu$}{B->X tau nu}}

\title{Exploring the \texorpdfstring{$\tau$}{tau} polarization in \texorpdfstring{$B\to X\tau\bar\nu$}{B->X tau nu} along different axes}

\author{Florian U.\ Bernlochner}
\affiliation{Physikalisches Institut der Rheinischen Friedrich-Wilhelms-Universit\"at Bonn, 53115 Bonn, Germany}

\author{Zoltan Ligeti}
\affiliation{Ernest Orlando Lawrence Berkeley National Laboratory, 
University of California, Berkeley, CA 94720, USA}
\affiliation{Berkeley Center for Theoretical Physics, 
Department of Physics,
University of California, Berkeley, CA 94720, USA}

\author{Michele Papucci}
\affiliation{Burke Institute for Theoretical Physics, 
California Institute of Technology, Pasadena, CA 91125, USA}

\author{Dean J.\ Robinson}
\affiliation{Ernest Orlando Lawrence Berkeley National Laboratory, 
University of California, Berkeley, CA 94720, USA}
\affiliation{Berkeley Center for Theoretical Physics, 
Department of Physics,
University of California, Berkeley, CA 94720, USA}

\begin{abstract}

The $\tau$ polarization in semileptonic $B$ decays provides probes of new physics complementary to decay rate distributions of the three-body final state.  
Prior calculations for inclusive decays used a definition for the polarization axis that is different from the choice used in calculations (and the only measurement) for exclusive channels.  
To compare inclusive and exclusive predictions, we calculate the $\tau$ polarization in inclusive $B\to X\tau\bar\nu$ using the same choice as in the exclusive decays,
and construct a sum rule relating the inclusive $\tau$ polarization to a weighted sum of exclusive decay polarizations.  
We use this relation, experimental data, and theoretical predictions for the decays to the lightest charm or up-type hadrons to make predictions for excited channels.

\end{abstract}

\maketitle

\section{Introduction}

Semileptonic $B$ decays to $\tau$ leptons have received immense attention over the last decade
because of tensions between BaBar, Belle, and LHCb measurements of ratios sensitive to lepton flavor universality (LFU) violation and the standard model (SM) expectations~\cite{HFLAV:2022pwe}.
For $b \to c \tau \bar\nu$ decays, the subsequent decay of the $\tau$ within the detector 
allows measurement of the $\tau$ polarization fraction, $P_\tau = [\Gamma(s_\tau = +) - \Gamma(s_\tau = -)]/\Gamma$, 
where $s_\tau$ is the~$\tau$~spin projection along a given polarization axis and $\Gamma$ is the total rate.
The $\tau$ polarization fraction (hereafter just `polarization') depends on the hadronic final state, and is sensitive to beyond SM contributions, 
providing a probe of new physics complementary to the branching ratios or differential distributions of the three-body final state 
(treating the $\tau$ as stable).

\begin{figure}[b]
\includegraphics[width=.45\columnwidth, clip, bb=0 5 90 85]{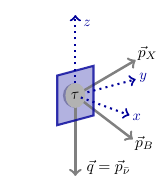}
\caption{The $B\to X\tau\bar\nu$ decay in the $\tau$ rest frame.
The three-momenta of the $B$, $X$, and $\bar\nu$ lie in the $yz$ plane.
Physical choices for the polarization axes are: 
(i)~$\vec p_{\bar\nu}$, used in most exclusive decays; (ii)~$\vec p_B$, used in past inclusive decay calculations; 
(iii)~the transverse direction, $x$, along which a nonzero polarization would violate $CP$; and (iv) the direction $\vec p_X$, which leaves a much-needed void in the literature.}
\label{fig:directions}
\end{figure}

The definition of the polarization depends on the choice of the polarization axis for $s_\tau$.
It has been conventional to define the $\tau$ polarization in inclusive $B \to X \tau \bar\nu$ decays, $P_\tau(X)$, 
by choosing the polarization axis to be the direction of the $\tau$ momentum in the $B$ rest frame, $\vec p_\tau/|\vec p_\tau|$~\cite{Kalinowski:1990ba, Falk:1994gw, Grossman:1994ax, Jezabek:1997rk, Ligeti:2021six}.
This is equivalent to choosing $-\vec{p}_B/|\vec{p}_B|$ in the $\tau$ rest frame,
and we therefore call this the $\vec{p}_B$ polarization axis (\pBPA) convention.
Figure~\ref{fig:directions} illustrates this choice for a generic
$B\to X\tau\bar\nu$ decay (for $X$ any hadronic system).
By contrast, prior exclusive calculations choose the polarization axis to be $\vec p_\tau/|\vec p_\tau|$ in the dilepton rest frame~\cite{Tanaka:1994ay, Tanaka:2010se, Datta:2012qk, Tanaka:2012nw}.
In this frame with this choice, the $\tau$ spin basis (anti)aligns with the neutrino helicity basis, 
leading to the simplification that in the SM the $s_\tau = +$ amplitude is exclusively proportional to the $\tau$ mass, $m_\tau$;
i.e., the $s_\tau = -$ amplitude contains no $m_\tau$-dependent terms.
This polarization axis choice is equivalent to  $-\vec{p}_{\bar\nu}/|\vec{p}_{\bar\nu}|$ in the $\tau$ rest frame 
(see also Fig.~\ref{fig:directions}),
and we therefore call this the $\vec{p}_{\bar\nu}$ polarization axis (\pNPA) convention.

The only polarization measurement to date was performed in $B \to D^*\tau\bar\nu$,
%(with sizable uncertainties), 
using single prong $\tau \to \pi \nu$ and $\tau \to \rho \nu$ decays,
and using the \pNPA convention to define the polarization,
$P_\tau(D^*) = -0.38 \pm 0.51 ^{+0.21}_{-0.16}$~\cite{Belle:2017ilt}.
As shown in Figure~\ref{fig:directions}, 
one could also define polarizations projecting the $\tau$ spin on the $\vec p_X$ direction, or the direction
transverse to the plane spanned by $\vec p_B$, $\vec p_X$, and $\vec p_{\bar\nu}$.
A nonzero $\tau$ polarization in this transverse direction, $x$, violates $CP$~\cite{Atwood:1993ka, Grossman:1994eb, Hwang:2015ica, Ivanov:2017mrj, Penalva:2021gef, Penalva:2021wye}.
It therefore vanishes in the SM, but could be generated by new physics.

In order to compare the prediction for inclusive $P_\tau(X)$ with the (weighted sum of) predictions for exclusive channels,
it is necessary to derive predictions for the $\tau$ polarization in $B \to X\tau\bar\nu$ in the \pNPA convention:
this is the purpose of this work.
We further show that one may construct a sum rule, which, when combined with experimental data for exclusive decays to the lightest charmed mesons in the final state, 
may be used to make predictions for the (average) $\tau$ polarization in excited channels.

\section{The inclusive calculation}

In the \pBPA convention, following the notation of Ref.~\cite{Falk:1994gw},
one decomposes the partial decay rates for $\tau$ spin projection ``up'' ($s_\tau=+$) or ``down'' ($s_\tau=-$) as
\beq\label{Bdef}
\Gamma\big(\Bbar\to X\,\tau(s_\tau=\pm)\,\bar\nu\big)=
    \frac12\, \Gamma \pm \widetilde\Gamma\,.
\eeq
The $\tau$ polarization in the \pBPA convention is then $P_\tau(X) = 2\widetilde\Gamma/ \Gamma$.
For the \pNPA convention, in order to distinguish from $\widetilde\Gamma$ in Eq.~\eqref{Bdef} we write instead
\beq\label{Wdef}
\Gamma\big(\Bbar\to X\,\tau(s_\tau=\pm)\,\bar\nu\big)=
    \frac12\, \Gamma\pm\widehat\Gamma\,.
\eeq
Then the polarization fraction becomes $P_\tau(X) = 2\widehat\Gamma/ \Gamma$.
(We emphasize that $s_\tau=\pm$ has different meanings in Eqs.~(\ref{Bdef}) and (\ref{Wdef}), as introduced above.)

We define the kinematic variables
\beq
\hqs= \frac{q^2}{m_b^2}\,,\qquad
y = \frac{2E_\tau}{m_b}\,,\qquad
x = \frac{2E_\nu}{m_b}\,,
\eeq
where $E_\tau$ and $E_\nu$ are the energies of the respective particles in the $B$ rest frame.
We also define the mass ratios
\beq
\rho = \frac{m_j^2}{m_b^2}\,,\qquad 
\hmts = \frac{m_\tau^2}{m_b^2}\,,
\eeq
where $j=c\,, u$.  Performing the OPE~\cite{Chay:1990da, Bigi:1993fe, Blok:1993va, Manohar:1993qn}, we find (for notations, see Ref.~\cite{Manohar:2000dt}),
\begin{widetext}
\begin{align}
\label{eqn:fulldiff}
\frac1{\Gamma_0}\, \frac{\d\widehat\Gamma}{\d\hqs \d y \d x}
&= 6\, \Theta \left(x-\frac{2(\hqs-\hmts)}{y+\sqrt{y^2-4\hmts}}\right)
  \Theta\left(\frac{2(\hqs-\hmts)}{y-\sqrt{y^2-4\hmts}}-x\right) \\ 
&\times \bigg\{(\hqs-\rt) (-2 W_1 + W_2 - y W_3 + \rt W_4) 
- x \big[ y W_2 - (\hqs + \rt) W_3 - 2\rt W_5 \big] + \frac{2 x^2 \rt}{\hqs-\rt} W_2 \bigg\}\,. \nn
\end{align}
Here $\Gamma_0 = (|V_{jb}|^2\, G_F^2\, m_b^5) / (192\pi^3)$.
Integrating over $x$ and $\hqs$ gives,
\begin{align}\label{dGhatdy}
\frac1{\Gamma_0}\, \frac{\d\widehat\Gamma}{\d y} &= \sqrt{y^2-4 \rt}\, \bigg\{
  3x_0^2\big[y^2 - 2y\, (1 + 3\rt) + 4\rt\,(2 + \rt)\big] + x_0^3\big[3y\,(1 +\rt)-y^2 - 8\rt  \big] 
  + 12 x_0^2\,  (1 + \rt - y)^2\, X \nn\\
& + \frac{\lambda_2\, x_0}{m_b^2(1 + \rt - y)}\, \bigg[
 12y(3 + 17\rt + 5\rt^2) - 30y^2(1 + 3\rt) + 15y^3 - 48\rt(4 + \rt) \nn\\
&\qquad + 3x_0 \big[ -2y(12 + 58\rt + 25\rt^2) + y^2(17 + 45\rt) - 5y^3 + 2\rt(55 + 21\rt + 10\rt^2) \big] \nn\\
&\qquad + 5x_0^2 \big[ 2y(3 + 7\rt) - 4y^2(1 + \rt) + y^3 - 4\rt(5 - \rt)\big] 
  + 12 (1 + \rt - y)^2 (1 + 5\rt - 5\rho)\, X
\bigg] \nn\\
& + \frac{\lambda_1}{3m_b^2(1 + \rt - y)^2}\, \bigg[
  - 24\rt (1 + 3\rt)y - 12\rt(1 + \rt) y^2 + 6(1 + 3\rt) y^3 - 3 y^4 + 48\rt^2 (2 + \rt)
\nn\\
&\qquad + 6x_0 \big[ -2\rt y (1 - 8\rt - \rt^2) - y^2(5 - 2\rt + 5\rt^2) + y^3(3 + \rt) - y^4 + 16\rt(1 - 2\rt) \big]
\nn\\
&\qquad + 3x_0^2 \big[ - 4\rt y(18 + 29\rt + 7\rt^2) + y^2(15 + 52\rt + 43\rt^2) - 8y^3(1 + 2\rt) + 2y^4 - 2\rt(7 - 70\rt - 9\rt^2 - 4\rt^3) \big]
\nn\\
&\qquad + 2 x_0^3 \big[ 40\rt y(1 + \rt) - 2y^2(5 + 11\rt + 5\rt^2) + 5y^3(1 + \rt) - y^4 + 2\rt(5 - 38\rt + 5\rt^2) \big] \nn\\
&\qquad + 12 (1 + \rt - y)^2 \big[ 2(1-\rho)^2 - 3 y (1 + \rt - \rho) + 2\rt(4+\rt - 2\rho) \big]\, X
\bigg] \bigg\} \,,
\end{align}
where $x_0 = 1-\rho/(1+\rt-y)$ as in Ref.~\cite{Falk:1994gw}, and
\beq
X = \frac{\rt}{\sqrt{y^2-4 \rt}}\, \ln \frac{y-2\rt + \sqrt{y^2-4\rt}}{y-2\rt - \sqrt{y^2-4\rt}}\,.
\eeq
For completeness we also derive the $\hqs$ dependence of the $\tau$ polarization is ($\d\Gamma/\d\hqs$ is given in Ref.~\cite{Ligeti:2014kia}),
\begin{align}\label{dGhatdq2}
\frac1{\Gamma_0}\, \frac{\d\widehat\Gamma}{\d \hqs} &= \bigg(\! 1 + \frac{\lambda_1+15\lambda_2}{2m_b^2} \bigg)
  \sqrt{(1+\rho-\hqs)^2-4\rho}\, \frac{(\hqs - \rt)^2}{\hat q^6} 
  \Big[ 2\hat q^6 - \hqs(1+\rho + \hqs)(1 + \rho + \rt) + 2\rt(1 - \rho)^2 + 4\rho\hqs \Big]
  \nn\\
& + \frac{6\lambda_2}{m_b^2} \frac{(\hqs - \rt)^2}{\hat q^6 \sqrt{(1+\rho-\hqs)^2-4\rho}} 
  \Big[ 2 \hat q^6 \rho - \hat q^4 [(1 - \rho)^2 + \rt(3 + \rho)] + (\hqs - 2\rt) (1 - \rho)^3 + \hqs \rt(1 - \rho) (5 + \rho) \Big] .
\end{align}
Integrating Eq.~\eqref{dGhatdy} over $y$ or Eq.~\eqref{dGhatdq2} over $\hqs$,
\begin{align}\label{Ghat}
\frac{\widehat\Gamma}{\Gamma_0} &= - \bigg( 1 + \frac{\lambda_1 + 3\lambda_2}{2m_b^2} \bigg) \bigg\{\frac{1}{6}\sqrt{\lambda}\, \Big[3(1+\rho)(1+\rho^2-8\rho) +47\rt(1-\rho+\rho^2) -5\rt\rho+11(1+\rho) \rt^2-\rt^3\Big]\nn\\*
&\qquad +4\rho^2\big[3+\rt(3\rt+2\rho-6)\big] \ln f_j
  -4\rt(1-\rho)\big[2(1-\rho)^2+3\rt(1+\rho)\big] \ln f_\tau \bigg\}\, \\*
& + \frac{\lambda_2}{m_b^2} \bigg\{\sqrt{\lambda}\, \Big[3(1-\rho)^3+\rt(1-\rho)(47\rho-5)+\rt^2(1-11\rho)+\rt^3\Big]
 -24\rt\rho \Big[2(1-\rho)^2-\rt(2-3\rho)\Big] \ln( f_j f_\tau)\bigg\}\,, \nn
\end{align}
\end{widetext}
where $f_{j,\tau}=x_{j,\tau}+\sqrt{x_{j,\tau}^2-1}$, $x_j=(1+\rho-\rt)/(2\sqrt{\rho})$, $x_\tau=(1+\rt-\rho)/(2\sqrt{\rt})$, and $\lambda=1-2(\rho+\rt)+(\rho-\rt)^2=4\rho(x_j^2-1)=4\rt(x_\tau^2-1)$.

The fact that $\lambda_1$ enters $\d\widehat\Gamma/\d q^2$ and $\widehat\Gamma$ in Eqs.~(\ref{dGhatdq2}) and (\ref{Ghat}) as $1 + \lambda_1/(2m_b^2)$ follows from  
reparametrization invariance~\cite{Luke:1992cs}.  
($\widetilde\Gamma$, however, does not have such a structure~\cite{Falk:1994gw}.)
The terms proportional to $\lambda_1$ can be obtained by ``averaging" over the
residual motion of the $b$ quark in the $B$ meson (i.e., writing $p_b = m_b v+k$ and averaging over $k$), which 
leaves $q$ unaffected~\cite{Manohar:1993qn}.  Therefore, $\vec s_\tau \cdot \vec p_\nu / |\vec p_\nu|$ (in the $\tau$ rest frame) is also unchanged, 
resulting in the $1 + \lambda_1/(2m_b^2)$ structure. 
At the same time,
$\vec s_\tau \cdot \vec p_B / |\vec p_B|$, which defines $\widetilde\Gamma$, is altered, and hence $\widetilde\Gamma$ does not have the simple $1 + \lambda_1/(2m_b^2)$ structure.

The limit of vanishing final-state quark mass, $B\to X_u\tau\bar\nu$, has additional interesting features, 
in that the $b$-quark distribution function in the $B$ meson plays an enhanced role compared to that in $B\to X_u\ell\bar\nu$~\cite{Ligeti:2021six}.  
This arises due to the combination of the facts that (i) the $b\to u$ semileptonic decay rate at maximal $E_\tau$ does not vanish at the free-quark decay level 
and (ii) the phase space is restricted because of the $\tau$ mass.

The $m_c\to0$ limit of Eq.~(\ref{dGhatdy}) generates singular distributions (i.e., terms containing $\delta(1+\rt-y)$ and its derivatives),
\begin{widetext}
\begin{align}\label{dGuhatdy}
\frac1{\Gamma_0}\, \frac{\d\widehat\Gamma_u}{\d y} &= \sqrt{y^2-4 \rt}\, \bigg\{
  - y(3-2y) + \rt(16-15y) + 12\rt^2 + 12 (1 + \rt - y)^2\, X \nn\\
&\qquad + \frac{\lambda_1}{3m_b^2}\, \Big[ - 5y^2 + \rt(74-24y) + 24\rt^2
+ 12 \big[2 - 3y + \rt(8 - 3y) + 2\rt^2\big]\, X \Big]\nn\\
&\qquad + \frac{\lambda_2}{m_b^2}\, \Big[ - y(6 + 5y) + \rt(38 - 30y) + 60\rt^2
  + 12(1 + 5\rt)(1 + \rt - y)\, X \Big] \bigg\}\, \theta(1+\rt-y) \nn\\
&  + \bigg[ \frac{\lambda_2}{2m_b^2}\, (11-5\rt) + \frac{\lambda_1}{6m_b^2}\, (1-11\rt)\bigg]\, (1-\rt)^3\, \delta(1+\rt-y) 
  + \frac{\lambda_1}{6m_b^2}\, (1 -\rt)^5\, \delta'(1+\rt-y) \,.
\end{align}
For completeness, the $\hqs$ distribution of the $\tau$ polarization is ($\d\Gamma_u/\d\hqs$ and $\d\widetilde\Gamma_u/\d\hqs$ are given in Ref.~\cite{Ligeti:2021six}),
\beq\label{dGuhatdq2}
\frac1{\Gamma_0}\, \frac{\d\widehat\Gamma_u}{\d \hqs} = - \bigg(1 + \frac{\lambda_1 + 3\lambda_2}{2m_b^2} \bigg) \frac{(1 - \hqs)^2}{\hat q^6}\, (\hqs - \rt)^2\, \Big[ \hqs (1 + 2 \hqs) - \rt (2 + \hqs) \Big]
  + \frac{6\lambda_2}{m_b^2}\, (\hqs - \rt)^2\, (3-2\hqs+\rt) \,.
\eeq
Integrating over $y$, or taking the $m_c\to 0$ limit of Eq.~\eqref{Ghat} gives,
\beq\label{Guhat}
\frac{\widehat\Gamma_u}{\Gamma_0} =  - \bigg(1 + \frac{\lambda_1 + 3\lambda_2}{2m_b^2} \bigg)\, 
  \bigg[ \frac12 + \frac{22\rt}3 - 6\rt^2 - 2\rt^3 + \frac{\rt^4}6 + 2\rt(2+3\rt) \ln\rt \bigg]
  + \frac{\lambda_2}{m_b^2}\, (1-\rt)^3\, (3+\rt) \,.
\eeq
\end{widetext}
This limit is smooth, unlike the $m_c\to 0$ limit of Eq.~\eqref{dGhatdy}.
For $\rt=0$, these results satisfy $-2\widehat\Gamma=\Gamma$, i.e., $P_\tau(X) = -1$, independent of the final state quark mass. 
This occurs because in the SM the leptons produced by the charged-current electroweak interaction are purely left handed in the massless limit. 

Since the $s_\tau = +$ amplitude is exclusively proportional to the lepton mass, $\d\widehat\Gamma/\d \hqs$ in Eqs.~(\ref{dGhatdq2}) and (\ref{dGuhatdq2}) obey
\beq
\frac2{(\hqs - \rt)^2}\, \frac{\d\widehat\Gamma}{\d \hqs} =
- \bigg[\frac1{(\hqs - \rt)^2}\, \frac{\d\Gamma}{\d \hqs} \bigg]_{\rt \to -\rt} \,.
\eeq
This relation holds in the SM to all orders.

In addition, angular momentum conservation in $B\to X_u\tau\bar\nu$ implies that the $\tau$ polarization is fully left handed at maximal $E_\tau$.  
The power-suppressed terms that enter at order $\lqcd^2/m_b^2$ also account for the nonperturbative shift of the $E_\tau$ endpoint from the parton level to the hadron level.  
As a result, the physical rate at maximal $E_\tau$ vanishes, although it is nonzero at the endpoint at the parton level.  
It was argued in Ref.~\cite{Ligeti:2021six} that only the most singular terms among the nonperturbative corrections need to satisfy $-2\, \d\widehat\Gamma_u = \d\Gamma_u$.  
Correspondingly, Eq.~\eqref{dGuhatdy} shows that the $\lambda_1\, \delta(1+\rt-y)$ term changes between the two conventions of the $\tau$ polarization fraction, $2\widetilde\Gamma/\Gamma$ and $2\widehat\Gamma/\Gamma$. 
However, the most singular $\lambda_1\, \delta'(1+\rt-y)$ and $\lambda_2\, \delta(1+\rt-y)$ terms are identical in $\d\widehat\Gamma_u/\d y$ and $\d\widetilde\Gamma_u/\d y$, 
and these terms are equal to $-1/2$ times the corresponding terms in $\d\Gamma_u/\d y$~\cite{Ligeti:2021six}.

The ${\cal O}(\alpha_s)$ perturbative corrections are known for the differential rate and the $\tau$ polarization in the \pBPA convention~\cite{Jezabek:1996db, Jezabek:1997rk},
but they have not been computed for the $\tau$ polarization defined in the \pNPA convention.  
We have not calculated the ${\cal O}(\alpha_s)$ perturbative corrections to $\widehat\Gamma$.
However, based on the results for $\Gamma$ and $\widetilde\Gamma$, we expect such ${\cal O}(\alpha_s)$ corrections to modify the polarization, 
$2\widehat\Gamma/\Gamma$, below the percent level (except very near the endpoints of the kinematic distributions).

%We have not calculated the ${\cal O}(\alpha_s)$ perturbative corrections to $\widehat\Gamma$:
%They are known for the differential rate and the $\tau$ polarization in the \pBPA convention~\cite{Jezabek:1996db, Jezabek:1997rk},
%but they have not been computed for the $\tau$ polarization defined in the \pNPA convention.  
%Based on the results for $\Gamma$ and $\widetilde\Gamma$, we expect such ${\cal O}(\alpha_s)$ corrections to modify the polarization, 
%$2\widehat\Gamma/\Gamma$, below the percent level (except very near the endpoints of the kinematic distributions).

We do not study in this paper endpoint regions of differential distributions of the $\tau$ polarization fraction.  
We expect, similar to the differential rates, that at fixed order in the operator product expansion (OPE) reliable predictions cannot be made very near maximal $q^2$ or $E_\tau$.  
Near maximal $E_\tau$ these effects are related to the $b$-quark distribution function in the $B$ meson (sometimes called the shape function).  
The OPE also breaks down near maximal $q^2$~\cite{Bauer:2000xf, Neubert:2000ch, Ligeti:2014kia} because the expansion parameter related to the energy release becomes small.
The upper limits of $q^2$ only differ at second order, by ${\cal O}(\lqcd^2)$, between the lowest order in the OPE, $(m_b - m_c)^2$, and the endpoint at the hadron level, $(m_B - m_D)^2$.
The lepton energy endpoint, however, is shifted at first order, by ${\cal O}(\lqcd)$.

\section{Numerical results and implications}

In the \pBPA polarization axis convention, $P_\tau(X_c) \simeq -0.71$~\cite{Falk:1994gw} and $P_\tau(X_u) \simeq -0.77$~\cite{Ligeti:2021six} for $B\to X_c\tau\bar\nu$ and $B\to X_u\tau\bar\nu$ decays, respectively.
Using $m_b = 4.7\,\GeV$, $m_c = 1.3\,\GeV$, $m_\tau = 1.777\,\GeV$, and expanding to linear order in $\lambda_{1,2}$, we find 
in the \pNPA convention
\begin{align}
P_\tau(X_c) &= 2\widehat\Gamma / \Gamma = -0.30 + 0.44\, \lambda_2 \approx -0.24 \,, \label{Ptauincl} \\
P_\tau(X_u) &= 2\widehat\Gamma_u / \Gamma_u = -0.40 + 0.33\, \lambda_2 \approx -0.36 \label{Ptauinclu} \,. 
\end{align}
Note that $\lambda_1$ drops out at this order, as it enters both $\widehat\Gamma$ and $\Gamma$ as $1+\lambda_1/(2m_b^2)$.  
Using $\lambda_2 = 0.12\,\GeV^2$, the corresponding second-order terms alter the polarization by nearly $18\%$ and $10\%$ in $B\to X_c\tau\bar\nu$ and $X_u\tau\bar\nu$, respectively,
compared to the lowest order contributions.
The reason is that the reduced phase space (due to $m_\tau$) enhances the importance of the $\lambda_2$ terms, and $P_\tau(X_c)$ and $P_\tau(X_u)$ have somewhat small values at lowest order.  
(Similar reasons led the authors of Ref.~\cite{Jezabek:1997rk} to consider the ${\cal O}(\alpha_s)$ corrections relative to $1-P_\tau$, 
which is an ${\cal O}(1)$ quantity everywhere in phase space, rather than $P_\tau$ itself.)  
Hence, these seemingly large corrections do not indicate that the OPE breaks down, 
and we estimate higher-order corrections to be smaller, impacting the results in Eqs.~\eqref{Ptauincl} and~\eqref{Ptauinclu} at or below the $0.02$ level.

In a recent fit of the form factors to $B\to D^{(*)}\ell\bar\nu$ data ($\ell = e$, $\mu$), Ref.~\cite{Bernlochner:2022ywh} obtained
\beq\label{Pdds}
 P_\tau(D) = 0.323 \pm 0.003\,, \qquad 
 P_\tau(D^*)= -0.494 \pm 0.005 \,,
\eeq
with a correlation of $\rho = 0.189$.
From the fit results of Refs~\cite{Bernlochner:2017jxt,Bernlochner:2016bci} we predict for the four $D^{**}$ states:
\begin{align}\label{Pdss}
    P_\tau(D_0^*) &=  0.10 \pm 0.02 \,, \qquad~~\, P_\tau(D_1^*) = -0.10 \pm 0.02 \,, \nn\\ 
    P_\tau(D_1) &=  -0.22 \pm 0.04 \,, \qquad P_\tau(D_2^*) = -0.33 \pm 0.04 \,.     
\end{align}

\begin{table}[tb]
\renewcommand*{\arraystretch}{1.2}
\newcolumntype{C}[1]{ >{\centering\arraybackslash $} m{#1} <{$}}
\newcolumntype{R}[1]{ >{\raggedright\arraybackslash $} m{#1} <{$}}
\newcolumntype{L}[1]{ >{\raggedleft\arraybackslash $} m{#1} <{$}}
\scalebox{0.86}{
\parbox{1.15\linewidth}{
\begin{tabular}{C{1.25cm}L{1.75cm}p{0.5cm}L{1.75cm}p{0.5cm}C{3cm}}
\hline\hline\\[-10pt]
\multirow{2}{*}{\text{hadron}} & \multicolumn{2}{c}{${\cal B}(\Bbar \to H_c \ell \nu)\,\text{(\%)}$} & \multicolumn{2}{c}{$R(H_c)$} & {\cal B}(\Bbar \to H_c \tau \nu)\,\text{(\%)} \\
& \multicolumn{2}{c}{(measured)} & \multicolumn{2}{c}{(prediction)} & \text{(prediction)}\\
\hline
D 		& 2.27 \pm 0.06 & \cite{HFLAV:2022pwe} & 0.288 \pm 0.04 & \cite{Bernlochner:2022ywh} & 0.65 \pm 0.02\\
D^* 		& 5.22 \pm 0.11 & \cite{HFLAV:2022pwe} & 0.249 \pm 0.03 & \cite{Bernlochner:2022ywh} & 1.30 \pm 0.03\\
D_0^	*		& 0.44 \pm 0.08 & \cite{Bernlochner:2016bci} & 0.08 \pm 0.03 & \cite{Bernlochner:2017jxt} & 0.032 \pm 0.017\\
D_1^\prime	& 0.20 \pm 0.05 & \cite{Bernlochner:2016bci} & 0.05 \pm 0.02 & \cite{Bernlochner:2017jxt} & 0.010 \pm 0.006\\
D_1^*		& 0.67 \pm 0.05 & \cite{Bernlochner:2016bci} & 0.10 \pm 0.02 & \cite{Bernlochner:2017jxt} & 0.064 \pm 0.008\\
D_2^*		& 0.30 \pm 0.04 & \cite{Bernlochner:2016bci} & 0.07 \pm 0.01 & \cite{Bernlochner:2017jxt} & 0.021 \pm 0.004\\
\hline
\!\!\sum D^{(*,**)} & \text{---} && \text{---} && 2.08 \pm 0.04 \\
\hline
 X_c 			& 10.65 \pm 0.16 & \cite{HFLAV:2022pwe} & 0.223\pm 0.005 & \cite{Ligeti:2014kia} & 2.37 \pm 0.06 \\
\hline\hline
\end{tabular}
}}
\caption{Isospin-averaged branching ratio measurements for light-lepton ($\ell = e$, $\mu$) semileptonic $B$ decays to the six lightest charmed mesons,
predictions for the corresponding SM LFU ratios, and the semitauonic branching fractions.}
\label{tab:BFs}
\end{table}

The inclusive polarization can be written as a weighted sum over exclusive polarization fractions,
yielding a sum rule
\begin{equation}\label{sumrule}
	P_\tau(X_c) = \sum_{H_c}\frac{\mathcal{B}(\Bbar \to H_c \tau \nu)\, P_\tau(H_c) }{\mathcal{B}(\Bbar \to X_c \tau \nu)} \,.
\end{equation}
The semitauonic branching fractions to $D^{(*)}$ and $D^{**}$ have not been precisely measured.
Therefore, we combine branching ratio measurements for the light-lepton semileptonic modes 
with SM predictions for the LFU ratios $R(H) = {\cal B}(B\to H\tau\bar\nu) /{\cal B}(B\to H \ell \bar\nu)$
to predict the semitauonic branching ratios.
For the exclusive modes, we use predictions from the same fits as in Eqs.~\eqref{Pdds} and~\eqref{Pdss},
hence within each heavy quark spin symmetry doublet, the two $R(H)$ and two $P_\tau(H)$ predictions are correlated.
These inputs and the predictions for the semitauonic branching ratios are shown in Table~\ref{tab:BFs}.
(For the inclusive prediction, 
using the different evaluations $R(X_c) = 0.221 \pm 0.004$~\cite{Rahimi:2022vlv} 
and/or ${\cal B}(\Bbar \to X_c \ell \bar \nu_\ell) = (10.48 \pm 0.13)\%$~\cite{Bernlochner:2022ucr}, result in slightly different predictions:
${\cal B}(\Bbar \to X_c \tau \bar \nu) = (2.34 \pm 0.06)\%$~\cite{Ligeti:2014kia,Bernlochner:2022ucr},
${\cal B}(\Bbar \to X_c \tau \bar \nu) = (2.32 \pm 0.06)\%$~\cite{Rahimi:2022vlv,Bernlochner:2022ucr},
${\cal B}(\Bbar \to X_c \tau \bar \nu) = (2.35 \pm 0.06)\%$~\cite{Rahimi:2022vlv,HFLAV:2022pwe}.)

The resulting contribution of the six lightest charm mesons in Eqs.~\eqref{Pdds}--\eqref{Pdss} to the inclusive polarization fraction is, 
\begin{align}\label{Psix}
P_\tau(D^{(*,**)}) & = \sum_{D, D^*, D^{**}}\frac{\mathcal{B}(\Bbar \to H_c \tau \nu)\, P_\tau(H_c) }{\mathcal{B}(\Bbar \to X_c \tau \nu)} \nonumber \\ &  = -0.190 \pm 0.007 \,.
\end{align}
Assuming that the remaining charm states, that saturate the inclusive $B\to X_c\tau\bar\nu$ width, 
all yield $\tau$ leptons with maximal (minimal) polarization, $P_\tau = +1$ ($-1$),
results in an upper (lower) bound for $P_\tau(X_c)$. 
One finds
\begin{align}\label{eq:Xclimit}
	P_\tau^{\rm min}(X_c) &= -0.31 \pm0.03\,, \nn\\*
 P_\tau^{\rm max}(X_c) &= -0.07\pm0.03 \,.
\end{align}
This is consistent with the prediction in Eq.~\eqref{Ptauincl}.

\begin{figure}[t]
\includegraphics[width=\columnwidth, clip, bb=15 45 585 390]{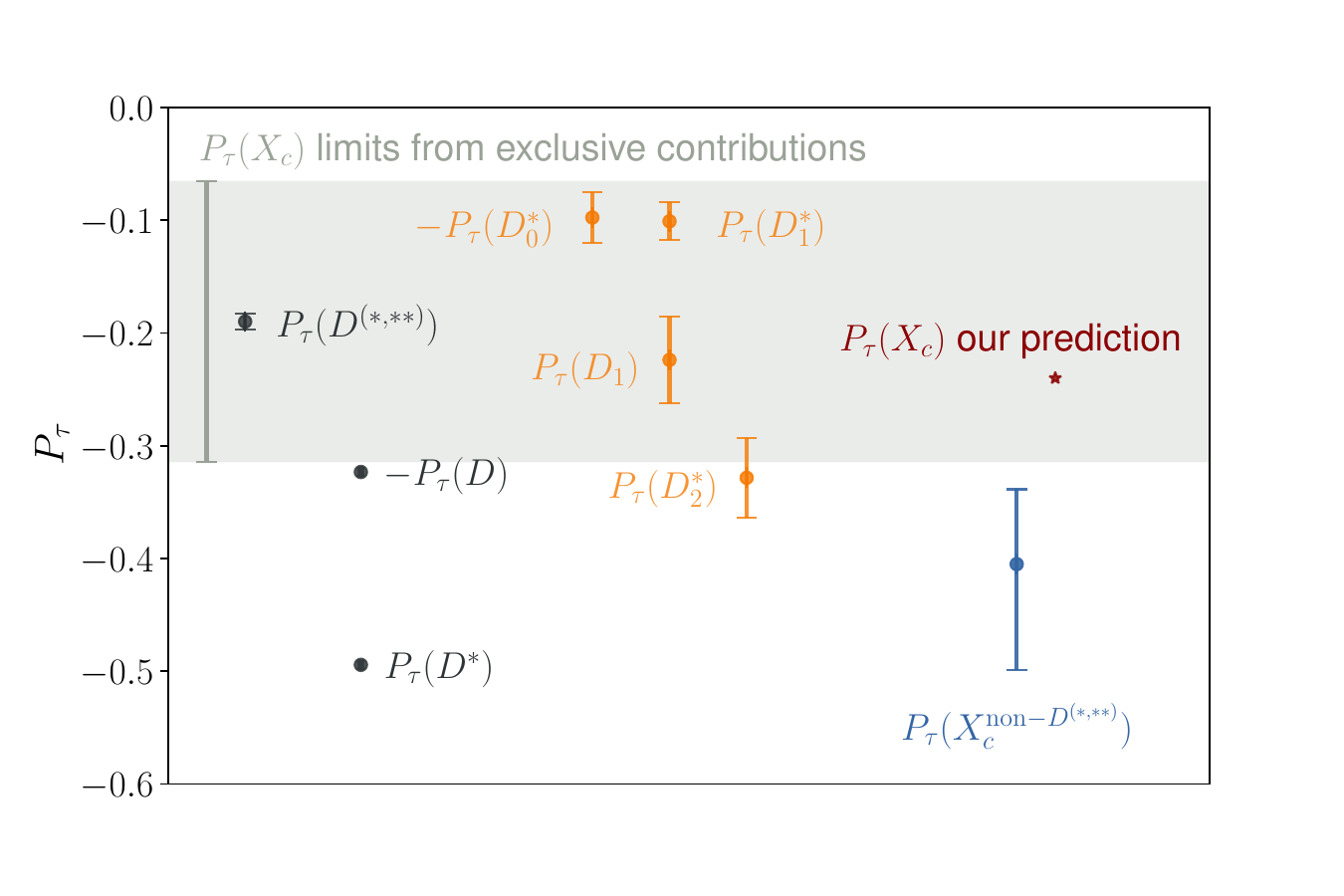}
\caption{Predictions for $P_\tau$ in the \pNPA convention.  The red point shows $P_\tau(X_c)$ in inclusive decay in Eq.~\eqref{Ptauincl}.  
The gray error bar and the shaded band shows the allowed range derived in Eq.~\eqref{eq:Xclimit}.  
The black error bars show predictions for the average of the six lightest states in Eq.~\eqref{Psix} and for $D^{(*)}$ in Eq.~\eqref{Pdds}.  
The orange error bars show predictions for $P_\tau(D^{**})$ in Eq.~\eqref{Pdss}.  
The blue error bar shows the predicted average polarization of the non-$D^{(*,**)}$ states in Eq.~\eqref{higherstates}.}\label{fig:PolSummary}
\end{figure}

Turning the sum rule in Eq.~\eqref{sumrule} around, we can use the inclusive polarization prediction in Eq.~\eqref{Ptauincl}
to predict the branching-ratio-weighted average polarization of higher excited charm states,
\begin{equation}\label{higherstates}
	P_\tau(X_c^{\mathrm{non-}D^{(*,**)}}) = -0.41^{+ 0.07}_{- 0.09} \,.
\end{equation}
Figure~\ref{fig:PolSummary} summarizes our predictions for $P_\tau$ in inclusive and exclusive decays
in the \pNPA convention.

Next, we consider the analog of the sum rule in Eq.~\eqref{sumrule} for $P_\tau(X_u)$.  
Predictions for the $\tau$ polarization and LFU ratios in exclusive charmless semitauonic decays to the lightest hadrons 
are available for $B \to \pi\tau\bar\nu$~\cite{Bernlochner:2015mya}, $\rho\tau\bar\nu$ and $\omega\tau\bar\nu$~\cite{Bernlochner:2021rel}. 
Using the latest BCL form factor parametrization from a combined fit to lattice QCD predictions plus BaBar and Belle data~\cite{FlavourLatticeAveragingGroupFLAG:2021npn},
one finds
\begin{equation}
	P_\tau(\pi) = -0.270 \pm 0.028\,,\qquad  R(\pi) = 0.653 \pm 0.015\,.
\end{equation}
(If instead one used the combined fit from Ref.~\cite{FermilabLattice:2015mwy}, one would find $P_\tau(\pi) = -0.296 \pm 0.029$ and $R(\pi) = 0.640 \pm 0.016$.)
A combined fit of averaged spectra from Belle and BaBar plus light-cone sum rule calculations yields~\cite{Bernlochner:2021rel}
\begin{align}
	P_\tau(\rho) & = -0.543 \pm 0.025\,, &  R(\rho) & = 0.532 \pm 0.011\,,  \nn\\
	P_\tau(\omega) & = -0.545 \pm 0.029\,, &  R(\omega) & = 0.534 \pm 0.018\,.
\end{align}
%(The signs of $P_\tau$ quoted here are opposite to Ref.~\cite{Bernlochner:2021rel}.)
Using in addition the prediction $R(X_u) = 0.337$~\cite{Hoang:1998hm, Ligeti:2021six} (no uncertainty is quoted)
we may derive bounds analogous to Eq.~\eqref{eq:Xclimit}. 
We find
\begin{align}\label{eq:Xulimit}
	P_\tau^{\rm min}(X_u^+) &= -0.72 \pm 0.04\,, \nn\\
    P_\tau^{\rm max}(X_u^+) &= 0.28 \pm 0.10\,,
\end{align}
which clearly satisfies Eq.~\eqref{Ptauinclu}.%
\footnote{One may instead obtain a lower bound for the semitauonic channels by assuming
\begin{equation}
	\label{eqn:taumu}
	\frac{{\cal B}(B\to H_u\tau\bar\nu)}{{\cal B}(B\to X_u\tau\bar\nu)} > \frac{{\cal B}(B\to H_u \mu \bar\nu)}{{\cal B}(B\to X_u \mu \bar\nu)} \,,
\end{equation}
motivated by the intuition that the reduction of the phase space due to the $\tau$ mass 
should enhance the fraction of the inclusive decay going into the lightest exclusive hadronic final states. 
This results in the looser bound $P_\tau^{\rm min} (X_u^+) = -0.84 \pm 0.02$.}
Here we used ${\cal B}(\Bbar^0 \to X_u^+ \ell \bar\nu) = (1.51 \pm 0.19)\times 10^{-3}$, 
${\cal B}(\Bbar^0 \to \pi^+ \ell \bar\nu) = (1.50 \pm 0.06)\times 10^{-4}$ and ${\cal B}(\Bbar^0 \to \rho^+ \ell \bar\nu) = (2.94 \pm 0.21)\times 10^{-4}$~\cite{Workman:2022ynf}.
The average polarization for higher excited light hadrons that would saturate Eq.~\eqref{Ptauinclu} is 
\begin{align}
P_\tau(X_u^{\mathrm{non-}\pi^+\!,\,\rho^+}) = -0.29^{+0.03}_{-0.02} \, .
\end{align}

\section{Summary}

We calculated the SM prediction for the $\tau$ polarization in inclusive semileptonic $B\to X\tau\bar\nu$ decay, 
choosing the \pNPA polarization axis convention to define $P_\tau$, in which the $\tau$ spin corresponds to the helicity in the $\tau\bar\nu$ rest frame.  
We derived differential distributions that may aid future measurements, and the total polarization is given in Eqs.~(\ref{Ptauincl}) and (\ref{Ptauinclu}).
These prediction were not previously available, and therefore comparisons between the polarization fractions in inclusive and exclusive decays could not be made.  
The sum rule in Eq.~\eqref{sumrule} relates the $\tau$ polarization fraction in inclusive decay to a branching-ratio-weighted sum over exclusive modes.  
We explored what is known about the SM predictions for the six lightest charm mesons ($D$, $D^{*}$, and $D^{**}$), 
which allowed us to make predictions for the average $\tau$ polarization in the remaining final states, that saturate the inclusive decay.
The similar analysis for charmless semileptonic $B$ decays is less constraining at present, but could prove useful with large data sets expected in the future.

\acknowledgments

We thank Aneesh Manohar for helpful discussion.
ZL thanks the Aspen Center for Physics (supported by the NSF Grant PHY-1607611) for hospitality while some of this work was carried out.
FB is supported by DFG Emmy-Noether Grant No.\ BE~6075/1-1 and BMBF Grant No.\ 05H21PDKBA. 
The work of ZL and DJR is supported by the Office of High Energy Physics of the U.S.\ Department of Energy under contract DE-AC02-05CH11231. 
MP is supported by the U.S.\ Department of Energy, Office of High Energy Physics, under Award Number DE-SC0011632 and by the Walter Burke Institute for Theoretical Physics.

% If not commented out, lists all entries in bib file
%\nocite*
\bibliography{refs}

\begin{thebibliography}{38}%
\makeatletter
\providecommand \@ifxundefined [1]{%
 \@ifx{#1\undefined}
}%
\providecommand \@ifnum [1]{%
 \ifnum #1\expandafter \@firstoftwo
 \else \expandafter \@secondoftwo
 \fi
}%
\providecommand \@ifx [1]{%
 \ifx #1\expandafter \@firstoftwo
 \else \expandafter \@secondoftwo
 \fi
}%
\providecommand \natexlab [1]{#1}%
\providecommand \enquote  [1]{``#1''}%
\providecommand \bibnamefont  [1]{#1}%
\providecommand \bibfnamefont [1]{#1}%
\providecommand \citenamefont [1]{#1}%
\providecommand \href@noop [0]{\@secondoftwo}%
\providecommand \href [0]{\begingroup \@sanitize@url \@href}%
\providecommand \@href[1]{\@@startlink{#1}\@@href}%
\providecommand \@@href[1]{\endgroup#1\@@endlink}%
\providecommand \@sanitize@url [0]{\catcode `\\12\catcode `\$12\catcode
  `\&12\catcode `\#12\catcode `\^12\catcode `\_12\catcode `\%12\relax}%
\providecommand \@@startlink[1]{}%
\providecommand \@@endlink[0]{}%
\providecommand \url  [0]{\begingroup\@sanitize@url \@url }%
\providecommand \@url [1]{\endgroup\@href {#1}{\urlprefix }}%
\providecommand \urlprefix  [0]{URL }%
\providecommand \Eprint [0]{\href }%
\providecommand \doibase [0]{http://dx.doi.org/}%
\providecommand \selectlanguage [0]{\@gobble}%
\providecommand \bibinfo  [0]{\@secondoftwo}%
\providecommand \bibfield  [0]{\@secondoftwo}%
\providecommand \translation [1]{[#1]}%
\providecommand \BibitemOpen [0]{}%
\providecommand \bibitemStop [0]{}%
\providecommand \bibitemNoStop [0]{.\EOS\space}%
\providecommand \EOS [0]{\spacefactor3000\relax}%
\providecommand \BibitemShut  [1]{\csname bibitem#1\endcsname}%
\let\auto@bib@innerbib\@empty
%</preamble>
\bibitem [{\citenamefont {Amhis}\ \emph {et~al.}(2022)\citenamefont {Amhis}
  \emph {et~al.}}]{HFLAV:2022pwe}%
  \BibitemOpen
  \bibfield  {author} {\bibinfo {author} {\bibfnamefont {Y.}~\bibnamefont
  {Amhis}} \emph {et~al.} (\bibinfo {collaboration} {HFLAV Collaboration}),\
  }\href@noop {} {\  (\bibinfo {year} {2022})},\ \Eprint
  {http://arxiv.org/abs/2206.07501} {arXiv:2206.07501 [hep-ex]} \BibitemShut
  {NoStop}%
\bibitem [{\citenamefont {Kalinowski}(1990)}]{Kalinowski:1990ba}%
  \BibitemOpen
  \bibfield  {author} {\bibinfo {author} {\bibfnamefont {J.}~\bibnamefont
  {Kalinowski}},\ }\href {\doibase 10.1016/0370-2693(90)90134-R} {\bibfield
  {journal} {\bibinfo  {journal} {Phys. Lett. B}\ }\textbf {\bibinfo {volume}
  {245}},\ \bibinfo {pages} {201} (\bibinfo {year} {1990})}\BibitemShut
  {NoStop}%
\bibitem [{\citenamefont {Falk}\ \emph {et~al.}(1994)\citenamefont {Falk},
  \citenamefont {Ligeti}, \citenamefont {Neubert},\ and\ \citenamefont
  {Nir}}]{Falk:1994gw}%
  \BibitemOpen
  \bibfield  {author} {\bibinfo {author} {\bibfnamefont {A.~F.}\ \bibnamefont
  {Falk}}, \bibinfo {author} {\bibfnamefont {Z.}~\bibnamefont {Ligeti}},
  \bibinfo {author} {\bibfnamefont {M.}~\bibnamefont {Neubert}}, \ and\
  \bibinfo {author} {\bibfnamefont {Y.}~\bibnamefont {Nir}},\ }\href {\doibase
  10.1016/0370-2693(94)91206-8} {\bibfield  {journal} {\bibinfo  {journal}
  {Phys. Lett. B}\ }\textbf {\bibinfo {volume} {326}},\ \bibinfo {pages} {145}
  (\bibinfo {year} {1994})},\ \Eprint {http://arxiv.org/abs/hep-ph/9401226}
  {arXiv:hep-ph/9401226} \BibitemShut {NoStop}%
\bibitem [{\citenamefont {Grossman}\ and\ \citenamefont
  {Ligeti}(1994)}]{Grossman:1994ax}%
  \BibitemOpen
  \bibfield  {author} {\bibinfo {author} {\bibfnamefont {Y.}~\bibnamefont
  {Grossman}}\ and\ \bibinfo {author} {\bibfnamefont {Z.}~\bibnamefont
  {Ligeti}},\ }\href {\doibase 10.1016/0370-2693(94)91267-X} {\bibfield
  {journal} {\bibinfo  {journal} {Phys. Lett. B}\ }\textbf {\bibinfo {volume}
  {332}},\ \bibinfo {pages} {373} (\bibinfo {year} {1994})},\ \Eprint
  {http://arxiv.org/abs/hep-ph/9403376} {arXiv:hep-ph/9403376} \BibitemShut
  {NoStop}%
\bibitem [{\citenamefont {Jezabek}\ and\ \citenamefont
  {Urban}(1998)}]{Jezabek:1997rk}%
  \BibitemOpen
  \bibfield  {author} {\bibinfo {author} {\bibfnamefont {M.}~\bibnamefont
  {Jezabek}}\ and\ \bibinfo {author} {\bibfnamefont {P.}~\bibnamefont
  {Urban}},\ }\href {\doibase 10.1016/S0550-3213(98)00189-8} {\bibfield
  {journal} {\bibinfo  {journal} {Nucl. Phys. B}\ }\textbf {\bibinfo {volume}
  {525}},\ \bibinfo {pages} {350} (\bibinfo {year} {1998})},\ \Eprint
  {http://arxiv.org/abs/hep-ph/9712440} {arXiv:hep-ph/9712440} \BibitemShut
  {NoStop}%
\bibitem [{\citenamefont {Ligeti}\ \emph {et~al.}(2022)\citenamefont {Ligeti},
  \citenamefont {Luke},\ and\ \citenamefont {Tackmann}}]{Ligeti:2021six}%
  \BibitemOpen
  \bibfield  {author} {\bibinfo {author} {\bibfnamefont {Z.}~\bibnamefont
  {Ligeti}}, \bibinfo {author} {\bibfnamefont {M.}~\bibnamefont {Luke}}, \ and\
  \bibinfo {author} {\bibfnamefont {F.~J.}\ \bibnamefont {Tackmann}},\ }\href
  {\doibase 10.1103/PhysRevD.105.073009} {\bibfield  {journal} {\bibinfo
  {journal} {Phys. Rev. D}\ }\textbf {\bibinfo {volume} {105}},\ \bibinfo
  {pages} {073009} (\bibinfo {year} {2022})},\ \Eprint
  {http://arxiv.org/abs/2112.07685} {arXiv:2112.07685 [hep-ph]} \BibitemShut
  {NoStop}%
\bibitem [{\citenamefont {Tanaka}(1995)}]{Tanaka:1994ay}%
  \BibitemOpen
  \bibfield  {author} {\bibinfo {author} {\bibfnamefont {M.}~\bibnamefont
  {Tanaka}},\ }\href {\doibase 10.1007/BF01571294} {\bibfield  {journal}
  {\bibinfo  {journal} {Z. Phys. C}\ }\textbf {\bibinfo {volume} {67}},\
  \bibinfo {pages} {321} (\bibinfo {year} {1995})},\ \Eprint
  {http://arxiv.org/abs/hep-ph/9411405} {arXiv:hep-ph/9411405} \BibitemShut
  {NoStop}%
\bibitem [{\citenamefont {Tanaka}\ and\ \citenamefont
  {Watanabe}(2010)}]{Tanaka:2010se}%
  \BibitemOpen
  \bibfield  {author} {\bibinfo {author} {\bibfnamefont {M.}~\bibnamefont
  {Tanaka}}\ and\ \bibinfo {author} {\bibfnamefont {R.}~\bibnamefont
  {Watanabe}},\ }\href {\doibase 10.1103/PhysRevD.82.034027} {\bibfield
  {journal} {\bibinfo  {journal} {Phys. Rev. D}\ }\textbf {\bibinfo {volume}
  {82}},\ \bibinfo {pages} {034027} (\bibinfo {year} {2010})},\ \Eprint
  {http://arxiv.org/abs/1005.4306} {arXiv:1005.4306 [hep-ph]} \BibitemShut
  {NoStop}%
\bibitem [{\citenamefont {Datta}\ \emph {et~al.}(2012)\citenamefont {Datta},
  \citenamefont {Duraisamy},\ and\ \citenamefont {Ghosh}}]{Datta:2012qk}%
  \BibitemOpen
  \bibfield  {author} {\bibinfo {author} {\bibfnamefont {A.}~\bibnamefont
  {Datta}}, \bibinfo {author} {\bibfnamefont {M.}~\bibnamefont {Duraisamy}}, \
  and\ \bibinfo {author} {\bibfnamefont {D.}~\bibnamefont {Ghosh}},\ }\href
  {\doibase 10.1103/PhysRevD.86.034027} {\bibfield  {journal} {\bibinfo
  {journal} {Phys. Rev. D}\ }\textbf {\bibinfo {volume} {86}},\ \bibinfo
  {pages} {034027} (\bibinfo {year} {2012})},\ \Eprint
  {http://arxiv.org/abs/1206.3760} {arXiv:1206.3760 [hep-ph]} \BibitemShut
  {NoStop}%
\bibitem [{\citenamefont {Tanaka}\ and\ \citenamefont
  {Watanabe}(2013)}]{Tanaka:2012nw}%
  \BibitemOpen
  \bibfield  {author} {\bibinfo {author} {\bibfnamefont {M.}~\bibnamefont
  {Tanaka}}\ and\ \bibinfo {author} {\bibfnamefont {R.}~\bibnamefont
  {Watanabe}},\ }\href {\doibase 10.1103/PhysRevD.87.034028} {\bibfield
  {journal} {\bibinfo  {journal} {Phys. Rev. D}\ }\textbf {\bibinfo {volume}
  {87}},\ \bibinfo {pages} {034028} (\bibinfo {year} {2013})},\ \Eprint
  {http://arxiv.org/abs/1212.1878} {arXiv:1212.1878 [hep-ph]} \BibitemShut
  {NoStop}%
\bibitem [{\citenamefont {Hirose}\ \emph {et~al.}(2018)\citenamefont {Hirose}
  \emph {et~al.}}]{Belle:2017ilt}%
  \BibitemOpen
  \bibfield  {author} {\bibinfo {author} {\bibfnamefont {S.}~\bibnamefont
  {Hirose}} \emph {et~al.} (\bibinfo {collaboration} {Belle Collaboration}),\
  }\href {\doibase 10.1103/PhysRevD.97.012004} {\bibfield  {journal} {\bibinfo
  {journal} {Phys. Rev. D}\ }\textbf {\bibinfo {volume} {97}},\ \bibinfo
  {pages} {012004} (\bibinfo {year} {2018})},\ \Eprint
  {http://arxiv.org/abs/1709.00129} {arXiv:1709.00129 [hep-ex]} \BibitemShut
  {NoStop}%
\bibitem [{\citenamefont {Atwood}\ \emph {et~al.}(1993)\citenamefont {Atwood},
  \citenamefont {Eilam},\ and\ \citenamefont {Soni}}]{Atwood:1993ka}%
  \BibitemOpen
  \bibfield  {author} {\bibinfo {author} {\bibfnamefont {D.}~\bibnamefont
  {Atwood}}, \bibinfo {author} {\bibfnamefont {G.}~\bibnamefont {Eilam}}, \
  and\ \bibinfo {author} {\bibfnamefont {A.}~\bibnamefont {Soni}},\ }\href
  {\doibase 10.1103/PhysRevLett.71.492} {\bibfield  {journal} {\bibinfo
  {journal} {Phys. Rev. Lett.}\ }\textbf {\bibinfo {volume} {71}},\ \bibinfo
  {pages} {492} (\bibinfo {year} {1993})},\ \Eprint
  {http://arxiv.org/abs/hep-ph/9303268} {arXiv:hep-ph/9303268} \BibitemShut
  {NoStop}%
\bibitem [{\citenamefont {Grossman}\ and\ \citenamefont
  {Ligeti}(1995)}]{Grossman:1994eb}%
  \BibitemOpen
  \bibfield  {author} {\bibinfo {author} {\bibfnamefont {Y.}~\bibnamefont
  {Grossman}}\ and\ \bibinfo {author} {\bibfnamefont {Z.}~\bibnamefont
  {Ligeti}},\ }\href {\doibase 10.1016/0370-2693(95)00070-2} {\bibfield
  {journal} {\bibinfo  {journal} {Phys. Lett. B}\ }\textbf {\bibinfo {volume}
  {347}},\ \bibinfo {pages} {399} (\bibinfo {year} {1995})},\ \Eprint
  {http://arxiv.org/abs/hep-ph/9409418} {arXiv:hep-ph/9409418} \BibitemShut
  {NoStop}%
\bibitem [{\citenamefont {Hwang}(2015)}]{Hwang:2015ica}%
  \BibitemOpen
  \bibfield  {author} {\bibinfo {author} {\bibfnamefont {D.~S.}\ \bibnamefont
  {Hwang}},\ }\href@noop {} {\  (\bibinfo {year} {2015})},\ \Eprint
  {http://arxiv.org/abs/1504.06933} {arXiv:1504.06933 [hep-ph]} \BibitemShut
  {NoStop}%
\bibitem [{\citenamefont {Ivanov}\ \emph {et~al.}(2017)\citenamefont {Ivanov},
  \citenamefont {K\"orner},\ and\ \citenamefont {Tran}}]{Ivanov:2017mrj}%
  \BibitemOpen
  \bibfield  {author} {\bibinfo {author} {\bibfnamefont {M.~A.}\ \bibnamefont
  {Ivanov}}, \bibinfo {author} {\bibfnamefont {J.~G.}\ \bibnamefont
  {K\"orner}}, \ and\ \bibinfo {author} {\bibfnamefont {C.-T.}\ \bibnamefont
  {Tran}},\ }\href {\doibase 10.1103/PhysRevD.95.036021} {\bibfield  {journal}
  {\bibinfo  {journal} {Phys. Rev. D}\ }\textbf {\bibinfo {volume} {95}},\
  \bibinfo {pages} {036021} (\bibinfo {year} {2017})},\ \Eprint
  {http://arxiv.org/abs/1701.02937} {arXiv:1701.02937 [hep-ph]} \BibitemShut
  {NoStop}%
\bibitem [{\citenamefont {Penalva}\ \emph
  {et~al.}(2021{\natexlab{a}})\citenamefont {Penalva}, \citenamefont
  {Hern\'andez},\ and\ \citenamefont {Nieves}}]{Penalva:2021gef}%
  \BibitemOpen
  \bibfield  {author} {\bibinfo {author} {\bibfnamefont {N.}~\bibnamefont
  {Penalva}}, \bibinfo {author} {\bibfnamefont {E.}~\bibnamefont
  {Hern\'andez}}, \ and\ \bibinfo {author} {\bibfnamefont {J.}~\bibnamefont
  {Nieves}},\ }\href {\doibase 10.1007/JHEP06(2021)118} {\bibfield  {journal}
  {\bibinfo  {journal} {JHEP}\ }\textbf {\bibinfo {volume} {06}},\ \bibinfo
  {pages} {118} (\bibinfo {year} {2021}{\natexlab{a}})},\ \Eprint
  {http://arxiv.org/abs/2103.01857} {arXiv:2103.01857 [hep-ph]} \BibitemShut
  {NoStop}%
\bibitem [{\citenamefont {Penalva}\ \emph
  {et~al.}(2021{\natexlab{b}})\citenamefont {Penalva}, \citenamefont
  {Hern\'andez},\ and\ \citenamefont {Nieves}}]{Penalva:2021wye}%
  \BibitemOpen
  \bibfield  {author} {\bibinfo {author} {\bibfnamefont {N.}~\bibnamefont
  {Penalva}}, \bibinfo {author} {\bibfnamefont {E.}~\bibnamefont
  {Hern\'andez}}, \ and\ \bibinfo {author} {\bibfnamefont {J.}~\bibnamefont
  {Nieves}},\ }\href {\doibase 10.1007/JHEP10(2021)122} {\bibfield  {journal}
  {\bibinfo  {journal} {JHEP}\ }\textbf {\bibinfo {volume} {10}},\ \bibinfo
  {pages} {122} (\bibinfo {year} {2021}{\natexlab{b}})},\ \Eprint
  {http://arxiv.org/abs/2107.13406} {arXiv:2107.13406 [hep-ph]} \BibitemShut
  {NoStop}%
\bibitem [{\citenamefont {Chay}\ \emph {et~al.}(1990)\citenamefont {Chay},
  \citenamefont {Georgi},\ and\ \citenamefont {Grinstein}}]{Chay:1990da}%
  \BibitemOpen
  \bibfield  {author} {\bibinfo {author} {\bibfnamefont {J.}~\bibnamefont
  {Chay}}, \bibinfo {author} {\bibfnamefont {H.}~\bibnamefont {Georgi}}, \ and\
  \bibinfo {author} {\bibfnamefont {B.}~\bibnamefont {Grinstein}},\ }\href
  {\doibase 10.1016/0370-2693(90)90916-T} {\bibfield  {journal} {\bibinfo
  {journal} {Phys. Lett. B}\ }\textbf {\bibinfo {volume} {247}},\ \bibinfo
  {pages} {399} (\bibinfo {year} {1990})}\BibitemShut {NoStop}%
\bibitem [{\citenamefont {Bigi}\ \emph {et~al.}(1993)\citenamefont {Bigi},
  \citenamefont {Shifman}, \citenamefont {Uraltsev},\ and\ \citenamefont
  {Vainshtein}}]{Bigi:1993fe}%
  \BibitemOpen
  \bibfield  {author} {\bibinfo {author} {\bibfnamefont {I.~I.~Y.}\
  \bibnamefont {Bigi}}, \bibinfo {author} {\bibfnamefont {M.~A.}\ \bibnamefont
  {Shifman}}, \bibinfo {author} {\bibfnamefont {N.~G.}\ \bibnamefont
  {Uraltsev}}, \ and\ \bibinfo {author} {\bibfnamefont {A.~I.}\ \bibnamefont
  {Vainshtein}},\ }\href {\doibase 10.1103/PhysRevLett.71.496} {\bibfield
  {journal} {\bibinfo  {journal} {Phys. Rev. Lett.}\ }\textbf {\bibinfo
  {volume} {71}},\ \bibinfo {pages} {496} (\bibinfo {year} {1993})},\ \Eprint
  {http://arxiv.org/abs/hep-ph/9304225} {arXiv:hep-ph/9304225} \BibitemShut
  {NoStop}%
\bibitem [{\citenamefont {Blok}\ \emph {et~al.}(1994)\citenamefont {Blok},
  \citenamefont {Koyrakh}, \citenamefont {Shifman},\ and\ \citenamefont
  {Vainshtein}}]{Blok:1993va}%
  \BibitemOpen
  \bibfield  {author} {\bibinfo {author} {\bibfnamefont {B.}~\bibnamefont
  {Blok}}, \bibinfo {author} {\bibfnamefont {L.}~\bibnamefont {Koyrakh}},
  \bibinfo {author} {\bibfnamefont {M.~A.}\ \bibnamefont {Shifman}}, \ and\
  \bibinfo {author} {\bibfnamefont {A.~I.}\ \bibnamefont {Vainshtein}},\ }\href
  {\doibase 10.1103/PhysRevD.50.3572} {\bibfield  {journal} {\bibinfo
  {journal} {Phys. Rev. D}\ }\textbf {\bibinfo {volume} {49}},\ \bibinfo
  {pages} {3356} (\bibinfo {year} {1994})},\ \bibinfo {note} {[Erratum:
  Phys.Rev.D 50, 3572 (1994)]},\ \Eprint {http://arxiv.org/abs/hep-ph/9307247}
  {arXiv:hep-ph/9307247} \BibitemShut {NoStop}%
\bibitem [{\citenamefont {Manohar}\ and\ \citenamefont
  {Wise}(1994)}]{Manohar:1993qn}%
  \BibitemOpen
  \bibfield  {author} {\bibinfo {author} {\bibfnamefont {A.~V.}\ \bibnamefont
  {Manohar}}\ and\ \bibinfo {author} {\bibfnamefont {M.~B.}\ \bibnamefont
  {Wise}},\ }\href {\doibase 10.1103/PhysRevD.49.1310} {\bibfield  {journal}
  {\bibinfo  {journal} {Phys. Rev. D}\ }\textbf {\bibinfo {volume} {49}},\
  \bibinfo {pages} {1310} (\bibinfo {year} {1994})},\ \Eprint
  {http://arxiv.org/abs/hep-ph/9308246} {arXiv:hep-ph/9308246} \BibitemShut
  {NoStop}%
\bibitem [{\citenamefont {Manohar}\ and\ \citenamefont
  {Wise}(2000)}]{Manohar:2000dt}%
  \BibitemOpen
  \bibfield  {author} {\bibinfo {author} {\bibfnamefont {A.~V.}\ \bibnamefont
  {Manohar}}\ and\ \bibinfo {author} {\bibfnamefont {M.~B.}\ \bibnamefont
  {Wise}},\ }\href {\doibase 10.1017/CBO9780511529351} {\emph {\bibinfo {title}
  {Heavy Quark Physics}}},\ Camb.\ Monogr.\ on Part.\ Phys., Nucl.\ Phys.,
  Cosmol.\ (\bibinfo  {publisher} {Cambridge University Press},\ \bibinfo
  {year} {2000})\BibitemShut {NoStop}%
\bibitem [{\citenamefont {Ligeti}\ and\ \citenamefont
  {Tackmann}(2014)}]{Ligeti:2014kia}%
  \BibitemOpen
  \bibfield  {author} {\bibinfo {author} {\bibfnamefont {Z.}~\bibnamefont
  {Ligeti}}\ and\ \bibinfo {author} {\bibfnamefont {F.~J.}\ \bibnamefont
  {Tackmann}},\ }\href {\doibase 10.1103/PhysRevD.90.034021} {\bibfield
  {journal} {\bibinfo  {journal} {Phys. Rev. D}\ }\textbf {\bibinfo {volume}
  {90}},\ \bibinfo {pages} {034021} (\bibinfo {year} {2014})},\ \Eprint
  {http://arxiv.org/abs/1406.7013} {arXiv:1406.7013 [hep-ph]} \BibitemShut
  {NoStop}%
\bibitem [{\citenamefont {Luke}\ and\ \citenamefont
  {Manohar}(1992)}]{Luke:1992cs}%
  \BibitemOpen
  \bibfield  {author} {\bibinfo {author} {\bibfnamefont {M.~E.}\ \bibnamefont
  {Luke}}\ and\ \bibinfo {author} {\bibfnamefont {A.~V.}\ \bibnamefont
  {Manohar}},\ }\href {\doibase 10.1016/0370-2693(92)91786-9} {\bibfield
  {journal} {\bibinfo  {journal} {Phys. Lett. B}\ }\textbf {\bibinfo {volume}
  {286}},\ \bibinfo {pages} {348} (\bibinfo {year} {1992})},\ \Eprint
  {http://arxiv.org/abs/hep-ph/9205228} {arXiv:hep-ph/9205228} \BibitemShut
  {NoStop}%
\bibitem [{\citenamefont {Jezabek}\ and\ \citenamefont
  {Motyka}(1997)}]{Jezabek:1996db}%
  \BibitemOpen
  \bibfield  {author} {\bibinfo {author} {\bibfnamefont {M.}~\bibnamefont
  {Jezabek}}\ and\ \bibinfo {author} {\bibfnamefont {L.}~\bibnamefont
  {Motyka}},\ }\href {\doibase 10.1016/S0550-3213(97)00341-6} {\bibfield
  {journal} {\bibinfo  {journal} {Nucl. Phys. B}\ }\textbf {\bibinfo {volume}
  {501}},\ \bibinfo {pages} {207} (\bibinfo {year} {1997})},\ \Eprint
  {http://arxiv.org/abs/hep-ph/9701358} {arXiv:hep-ph/9701358} \BibitemShut
  {NoStop}%
\bibitem [{\citenamefont {Bauer}\ \emph {et~al.}(2000)\citenamefont {Bauer},
  \citenamefont {Ligeti},\ and\ \citenamefont {Luke}}]{Bauer:2000xf}%
  \BibitemOpen
  \bibfield  {author} {\bibinfo {author} {\bibfnamefont {C.~W.}\ \bibnamefont
  {Bauer}}, \bibinfo {author} {\bibfnamefont {Z.}~\bibnamefont {Ligeti}}, \
  and\ \bibinfo {author} {\bibfnamefont {M.~E.}\ \bibnamefont {Luke}},\ }\href
  {\doibase 10.1016/S0370-2693(00)00318-X} {\bibfield  {journal} {\bibinfo
  {journal} {Phys. Lett. B}\ }\textbf {\bibinfo {volume} {479}},\ \bibinfo
  {pages} {395} (\bibinfo {year} {2000})},\ \Eprint
  {http://arxiv.org/abs/hep-ph/0002161} {arXiv:hep-ph/0002161} \BibitemShut
  {NoStop}%
\bibitem [{\citenamefont {Neubert}(2000)}]{Neubert:2000ch}%
  \BibitemOpen
  \bibfield  {author} {\bibinfo {author} {\bibfnamefont {M.}~\bibnamefont
  {Neubert}},\ }\href {\doibase 10.1088/1126-6708/2000/07/022} {\bibfield
  {journal} {\bibinfo  {journal} {JHEP}\ }\textbf {\bibinfo {volume} {07}},\
  \bibinfo {pages} {022} (\bibinfo {year} {2000})},\ \Eprint
  {http://arxiv.org/abs/hep-ph/0006068} {arXiv:hep-ph/0006068} \BibitemShut
  {NoStop}%
\bibitem [{\citenamefont {Bernlochner}\ \emph
  {et~al.}(2022{\natexlab{a}})\citenamefont {Bernlochner}, \citenamefont
  {Ligeti}, \citenamefont {Papucci}, \citenamefont {Prim}, \citenamefont
  {Robinson},\ and\ \citenamefont {Xiong}}]{Bernlochner:2022ywh}%
  \BibitemOpen
  \bibfield  {author} {\bibinfo {author} {\bibfnamefont {F.~U.}\ \bibnamefont
  {Bernlochner}}, \bibinfo {author} {\bibfnamefont {Z.}~\bibnamefont {Ligeti}},
  \bibinfo {author} {\bibfnamefont {M.}~\bibnamefont {Papucci}}, \bibinfo
  {author} {\bibfnamefont {M.~T.}\ \bibnamefont {Prim}}, \bibinfo {author}
  {\bibfnamefont {D.~J.}\ \bibnamefont {Robinson}}, \ and\ \bibinfo {author}
  {\bibfnamefont {C.}~\bibnamefont {Xiong}},\ }\href {\doibase
  10.1103/PhysRevD.106.096015} {\bibfield  {journal} {\bibinfo  {journal}
  {Phys. Rev. D}\ }\textbf {\bibinfo {volume} {106}},\ \bibinfo {pages}
  {096015} (\bibinfo {year} {2022}{\natexlab{a}})},\ \Eprint
  {http://arxiv.org/abs/2206.11281} {arXiv:2206.11281 [hep-ph]} \BibitemShut
  {NoStop}%
\bibitem [{\citenamefont {Bernlochner}\ \emph {et~al.}(2018)\citenamefont
  {Bernlochner}, \citenamefont {Ligeti},\ and\ \citenamefont
  {Robinson}}]{Bernlochner:2017jxt}%
  \BibitemOpen
  \bibfield  {author} {\bibinfo {author} {\bibfnamefont {F.~U.}\ \bibnamefont
  {Bernlochner}}, \bibinfo {author} {\bibfnamefont {Z.}~\bibnamefont {Ligeti}},
  \ and\ \bibinfo {author} {\bibfnamefont {D.~J.}\ \bibnamefont {Robinson}},\
  }\href {\doibase 10.1103/PhysRevD.97.075011} {\bibfield  {journal} {\bibinfo
  {journal} {Phys. Rev. D}\ }\textbf {\bibinfo {volume} {97}},\ \bibinfo
  {pages} {075011} (\bibinfo {year} {2018})},\ \Eprint
  {http://arxiv.org/abs/1711.03110} {arXiv:1711.03110 [hep-ph]} \BibitemShut
  {NoStop}%
\bibitem [{\citenamefont {Bernlochner}\ and\ \citenamefont
  {Ligeti}(2017)}]{Bernlochner:2016bci}%
  \BibitemOpen
  \bibfield  {author} {\bibinfo {author} {\bibfnamefont {F.~U.}\ \bibnamefont
  {Bernlochner}}\ and\ \bibinfo {author} {\bibfnamefont {Z.}~\bibnamefont
  {Ligeti}},\ }\href {\doibase 10.1103/PhysRevD.95.014022} {\bibfield
  {journal} {\bibinfo  {journal} {Phys. Rev. D}\ }\textbf {\bibinfo {volume}
  {95}},\ \bibinfo {pages} {014022} (\bibinfo {year} {2017})},\ \Eprint
  {http://arxiv.org/abs/1606.09300} {arXiv:1606.09300 [hep-ph]} \BibitemShut
  {NoStop}%
\bibitem [{\citenamefont {Rahimi}\ and\ \citenamefont
  {Vos}(2022)}]{Rahimi:2022vlv}%
  \BibitemOpen
  \bibfield  {author} {\bibinfo {author} {\bibfnamefont {M.}~\bibnamefont
  {Rahimi}}\ and\ \bibinfo {author} {\bibfnamefont {K.~K.}\ \bibnamefont
  {Vos}},\ }\href {\doibase 10.1007/JHEP11(2022)007} {\bibfield  {journal}
  {\bibinfo  {journal} {JHEP}\ }\textbf {\bibinfo {volume} {11}},\ \bibinfo
  {pages} {007} (\bibinfo {year} {2022})},\ \Eprint
  {http://arxiv.org/abs/2207.03432} {arXiv:2207.03432 [hep-ph]} \BibitemShut
  {NoStop}%
\bibitem [{\citenamefont {Bernlochner}\ \emph
  {et~al.}(2022{\natexlab{b}})\citenamefont {Bernlochner}, \citenamefont
  {Fael}, \citenamefont {Olschewsky}, \citenamefont {Persson}, \citenamefont
  {van Tonder}, \citenamefont {Vos},\ and\ \citenamefont
  {Welsch}}]{Bernlochner:2022ucr}%
  \BibitemOpen
  \bibfield  {author} {\bibinfo {author} {\bibfnamefont {F.}~\bibnamefont
  {Bernlochner}}, \bibinfo {author} {\bibfnamefont {M.}~\bibnamefont {Fael}},
  \bibinfo {author} {\bibfnamefont {K.}~\bibnamefont {Olschewsky}}, \bibinfo
  {author} {\bibfnamefont {E.}~\bibnamefont {Persson}}, \bibinfo {author}
  {\bibfnamefont {R.}~\bibnamefont {van Tonder}}, \bibinfo {author}
  {\bibfnamefont {K.~K.}\ \bibnamefont {Vos}}, \ and\ \bibinfo {author}
  {\bibfnamefont {M.}~\bibnamefont {Welsch}},\ }\href {\doibase
  10.1007/JHEP10(2022)068} {\bibfield  {journal} {\bibinfo  {journal} {JHEP}\
  }\textbf {\bibinfo {volume} {10}},\ \bibinfo {pages} {068} (\bibinfo {year}
  {2022}{\natexlab{b}})},\ \Eprint {http://arxiv.org/abs/2205.10274}
  {arXiv:2205.10274 [hep-ph]} \BibitemShut {NoStop}%
\bibitem [{\citenamefont {Bernlochner}(2015)}]{Bernlochner:2015mya}%
  \BibitemOpen
  \bibfield  {author} {\bibinfo {author} {\bibfnamefont {F.~U.}\ \bibnamefont
  {Bernlochner}},\ }\href {\doibase 10.1103/PhysRevD.92.115019} {\bibfield
  {journal} {\bibinfo  {journal} {Phys. Rev. D}\ }\textbf {\bibinfo {volume}
  {92}},\ \bibinfo {pages} {115019} (\bibinfo {year} {2015})},\ \Eprint
  {http://arxiv.org/abs/1509.06938} {arXiv:1509.06938 [hep-ph]} \BibitemShut
  {NoStop}%
\bibitem [{\citenamefont {Bernlochner}\ \emph {et~al.}(2021)\citenamefont
  {Bernlochner}, \citenamefont {Prim},\ and\ \citenamefont
  {Robinson}}]{Bernlochner:2021rel}%
  \BibitemOpen
  \bibfield  {author} {\bibinfo {author} {\bibfnamefont {F.~U.}\ \bibnamefont
  {Bernlochner}}, \bibinfo {author} {\bibfnamefont {M.~T.}\ \bibnamefont
  {Prim}}, \ and\ \bibinfo {author} {\bibfnamefont {D.~J.}\ \bibnamefont
  {Robinson}},\ }\href {\doibase 10.1103/PhysRevD.104.034032} {\bibfield
  {journal} {\bibinfo  {journal} {Phys. Rev. D}\ }\textbf {\bibinfo {volume}
  {104}},\ \bibinfo {pages} {034032} (\bibinfo {year} {2021})},\ \Eprint
  {http://arxiv.org/abs/2104.05739} {arXiv:2104.05739 [hep-ph]} \BibitemShut
  {NoStop}%
\bibitem [{\citenamefont {Aoki}\ \emph {et~al.}(2022)\citenamefont {Aoki} \emph
  {et~al.}}]{FlavourLatticeAveragingGroupFLAG:2021npn}%
  \BibitemOpen
  \bibfield  {author} {\bibinfo {author} {\bibfnamefont {Y.}~\bibnamefont
  {Aoki}} \emph {et~al.} (\bibinfo {collaboration} {Flavour Lattice Averaging
  Group (FLAG)}),\ }\href {\doibase 10.1140/epjc/s10052-022-10536-1} {\bibfield
   {journal} {\bibinfo  {journal} {Eur. Phys. J. C}\ }\textbf {\bibinfo
  {volume} {82}},\ \bibinfo {pages} {869} (\bibinfo {year} {2022})},\ \Eprint
  {http://arxiv.org/abs/2111.09849} {arXiv:2111.09849 [hep-lat]} \BibitemShut
  {NoStop}%
\bibitem [{\citenamefont {Bailey}\ \emph {et~al.}(2015)\citenamefont {Bailey}
  \emph {et~al.}}]{FermilabLattice:2015mwy}%
  \BibitemOpen
  \bibfield  {author} {\bibinfo {author} {\bibfnamefont {J.~A.}\ \bibnamefont
  {Bailey}} \emph {et~al.} (\bibinfo {collaboration} {Fermilab Lattice,
  MILC}),\ }\href {\doibase 10.1103/PhysRevD.92.014024} {\bibfield  {journal}
  {\bibinfo  {journal} {Phys. Rev. D}\ }\textbf {\bibinfo {volume} {92}},\
  \bibinfo {pages} {014024} (\bibinfo {year} {2015})},\ \Eprint
  {http://arxiv.org/abs/1503.07839} {arXiv:1503.07839 [hep-lat]} \BibitemShut
  {NoStop}%
\bibitem [{\citenamefont {Hoang}\ \emph {et~al.}(1999)\citenamefont {Hoang},
  \citenamefont {Ligeti},\ and\ \citenamefont {Manohar}}]{Hoang:1998hm}%
  \BibitemOpen
  \bibfield  {author} {\bibinfo {author} {\bibfnamefont {A.~H.}\ \bibnamefont
  {Hoang}}, \bibinfo {author} {\bibfnamefont {Z.}~\bibnamefont {Ligeti}}, \
  and\ \bibinfo {author} {\bibfnamefont {A.~V.}\ \bibnamefont {Manohar}},\
  }\href {\doibase 10.1103/PhysRevD.59.074017} {\bibfield  {journal} {\bibinfo
  {journal} {Phys. Rev. D}\ }\textbf {\bibinfo {volume} {59}},\ \bibinfo
  {pages} {074017} (\bibinfo {year} {1999})},\ \Eprint
  {http://arxiv.org/abs/hep-ph/9811239} {arXiv:hep-ph/9811239} \BibitemShut
  {NoStop}%
\bibitem [{\citenamefont {Workman}\ \emph {et~al.}(2022)\citenamefont {Workman}
  \emph {et~al.}}]{Workman:2022ynf}%
  \BibitemOpen
  \bibfield  {author} {\bibinfo {author} {\bibfnamefont {R.~L.}\ \bibnamefont
  {Workman}} \emph {et~al.} (\bibinfo {collaboration} {Particle Data Group}),\
  }\href {\doibase 10.1093/ptep/ptac097} {\bibfield  {journal} {\bibinfo
  {journal} {PTEP}\ }\textbf {\bibinfo {volume} {2022}},\ \bibinfo {pages}
  {083C01} (\bibinfo {year} {2022})}\BibitemShut {NoStop}%
\end{thebibliography}%

\end{document}